\documentclass[12pt,preprint]{aastex}

\slugcomment{submitted to ApJ}

\shorttitle{LMC Bump Cepheids}
\shortauthors{Keller and Wood}

\begin{document}


\title{Large Magellanic Cloud Bump Cepheids: Probing the Stellar Mass-Luminosity Relation }


\author{S.\ C.\ Keller}
\affil{IGPP, L-413, LLNL, PO Box 505, Livermore, CA 94550}
\email{skeller@igpp.ucllnl.org}

\and

\author{P.\ R.\ Wood}
\affil{RSAA, Australian National University, Canberra A.C.T.~2600, Australia}
\email{wood@mso.anu.edu.au}



\begin{abstract}

We present the results of non-linear pulsation modelling of 20 bump
Cepheids in the LMC. By obtaining a optimal fit to the observed $V,R$
MACHO lightcurves we have placed tight constraints on stellar
parameters of $M$, $L$, $T_{eff}$ and well as quantities of distance
modulus and reddening. We describe the mass-luminosity relation for
core-He burning for intermediate mass stars. The mass-luminosity
relation depends critically on the level of mixing within the stellar
interior over the course of the main-sequence lifetime. Our sample is
significantly more luminous than predicted by classical stellar
evolutionary models that do not incorporate extension to the
convective core. Under the paradigm of convective core overshoot our
data implies $\Lambda_c$ of 0.65$\pm$0.03$l/H_p$. We derive a LMC
distance modulus of 18.55$\pm$0.02.
\end{abstract}


\keywords{Cepheids:pulsation stellar:evolution}


\section{Introduction}

Cepheids are classical distance indicators. Their tight conformity
to a period-luminosity relationship has made them the fundamental
basis of the extra-galactic distance scale and hence integral to
observational cosmology. Ideally, we would like to have theoretical
models capable of accurately predicting the period-luminosity relation
and its metallicity dependence. The regularity of Cepheid pulsation
provides a set of well defined observational parameters with which to
confront the predictions of theoretical models of stellar
pulsations. In this way, Cepheids provide us the ability to closely
scrutinise the accuracy of input physics within pulsation models.

Cepheid light curves display a variety of shapes and amplitudes that
are period dependent (the \citet{hea26} progression). A
feature of the lightcurves of some Cepheids is a pronounced bump
either preceding or following maximum. The bump arises from resonance
between the fundamental mode and second overtone. This resonance
becomes particularly prominent when the period ratio of these two
modes ($P_0/P_2$=$P_{02}$) is $\sim 2$.

The bump enables us to break the degeneracy that exists between
observable quantities; lightcurve shape, amplitude, period and the
intrinsic properties; mass, luminosity and temperature of the
Cepheid. Such a technique was first proposed by \citet{sto69}
and was demonstrated by \citet{woo97}(hereafter Paper 1) in their
non-linear pulsation analysis of the LMC bump Cepheid HV 905.

Through the analysis of bump Cepheids we have a probe of the stellar
mass-luminosity (M-L) relation for core He-burning stars. The M-L
relation depends critically on the size of the central He core
established (largely) during the course of the star's main-sequence
lifetime. The size of the He core is in turn, determined by the extent
of convection in the vicinity of the convective core.

The treatment of convection remains the weakest point in our
description of massive stars. Ongoing debate focuses on the degree of
extension to the convective core beyond that predicted by standard,
non-rotating stellar evolution models. Extension of the convective
core has traditionally been discussed in terms of convective core
overshoot (CCO) in the formalism of mixing-length theory. The CCO
parameter, $\Lambda_c$, sets the height (as a fraction of the pressure
scale height) to which gas packets from the convective core rise into
the formally convectively stable region surrounding the core.

Mixing in the vicinity of the convective core produces a number of
important evolutionary changes that are expressed in a stellar
population. It expands the amount of H available to the core and hence
extends the main-sequence lifetime. The star consequently develops a
more massive He core and the subsequent post-main-sequence evolution
occurs at a more rapid pace and at higher luminosities. That is, the
M-L relation is significantly more luminous than that of classical models.

Numerous studies have attempted to ascertain the efficiency of CCO
from a theoretical basis with results that range from negligible to
substantial (see e.g.~\citet{bre81}). An analytical approach appears
limited given the complexity of the phenomenon. Laboratory fluid
dynamics shows that an understanding of convective mixing requires a
description of the turbulence field at all scales, a problem that will
require detailed hydrodynamical modelling.

Observations are required to ascertain the amount of CCO to apply in
stellar evolutionary models. Many studies have sought to do so through
the study of young cluster populations (most recently
\citet{bar02,kel01}) and from the broader field population
\citep{bea01,cor02} of the Magellanic Clouds. Whilst large
uncertainties exist in the derived values of $\Lambda_c$, the broad
consensus of these studies is the necessity of some level of
CCO. 

Another way of quantifying $\Lambda_c$ is to use masses and
luminosities of Cepheids. Dynamical masses for Cepheids are presented
by \citet{eva97,eva98} and \citet{boh97a,boh97b}. Derived masses have
considerable uncertainties but the combined sample \citep{eva98}
indicates the necessity for some level of CCO.

This study aims to quantitatively establish the level of CCO by an
examination of the M-L relation of a sample of bump Cepheids from the
LMC. A consistent result of pulsation modelling is that the M-L
relation for Cepheids is significantly brighter than predicted by
classical stellar evolution. The study of the bump Cepheid HV905 in
Paper 1 found a bump mass 29\% lower than that required by
evolutionary models without CCO. Recently, \citet{bon02} applied
non-linear modelling techniques that incorporate turbulent convection
to two LMC bump Cepheids. They found that an acceptable match between
model and observed lightcurves required a mass-luminosity relation in
which stars are $\sim$15\% lower in mass than predicted by
evolutionary models that neglect convective core overshoot. Linear
pulsation analyses \citep{seb95,kan94} similarly require pulsation
masses for Cepheids that are significantly lower than evolution
masses.

\section{Observations}

Photometry for the LMC bump Cepheids considered here is taken from the
MACHO photometric database. Stars are only considered from the
central bar region (the top 22 MACHO fields) in which
standardised photometry exists. Magnitudes in the MACHO B and R
passbands have been converted to Kron-Cousins $V$ and $R$ using
existing transofromations described in \citet{alc99}. Photometric
uncertainties are quoted as $\pm0.035$ mag in zero point and $V$$-$$R$
colour. The observed Cepheids are listed in Table
\ref{tbl-1}.

\section{Model Details}

Details of the non-linear pulsation code are given in Paper 1.  The
opacities have been updated to OPAL~96 \citep{igl96}, supplemented at
low temperatures by those of \citet{ale94}. Convective energy
transport was included by means of mixing-length theory with the
assumption of a mixing length of 1.6 pressure scale heights. A linear
non-adiabatic code was used to derive the starting model for each
simulation. Models contained 460 mass points outside an inner radius
of 0.3 R$_{\odot}$. Transformation of $L$ and $T_{eff}$ into $V$ and
$V$$-$$R$ of our observations was made through interpolation into a
grid of synthetically derived colours and bolometric corrections. The
colors were computed for the revised \citet{kur93} fluxes used in
\citet{bes98} and described in more detail in
\citet{cas99}. Magnitudes were computed through energy integration
using the passbands of \citet{bes90}. We computed non-linear models at
a fixed composition of Y=0.27 and Z=0.008 found for young objects in
the LMC \citep{rus89}.

In contrast to \citet{bon02}, our method uses only stellar
pulsation and stellar atmosphere theory, we do not make recourse to
existing mass-luminosity (M-L) relations. Once abundance is assumed
three parameters, $M$, $L$ and $T_{eff}$ remain to characterise the
the pulsating envelope of each Cepheid. Thus three conditions are
required to determine these quantities.

The first condition is that the fundamental pulsation period of the
starting linear models must satisfy the observed period of the
Cepheid. The other two conditions come from fitting non-linear model
lightcurves to the observations. The two parameters we chose as
independent variables for the lightcurve fit were $T_{eff}$ and
P$_{02}$, the ratio of the fundamental to second overtone periods. The
amplitude of pulsation is dependent on the star's temperature relative
to the blue edge of the IS. Hence the amplitude of pulsation is a
strong constraint on $T_{eff}$. The phase of the bump is dependent on
P$_{02}$ and furthermore, is independent of the pulsation amplitude
\citep{sim81}.

To commence the modelling process values of $T_{eff}$ and P$_{02}$ were
specified and parameters $L$ and $M$ were iterated until the required
linear period and P$_{02}$ were produced. Once model parameters were
determined, the static model was perturbed with the eigenfunction of the
linear adiabatic fundamental mode. The perturbed model was run until
the kinetic energy of the pulsation reached a limit cycle.

Our models incorporate convective energy transfer through the mixing
length approximation. This is known to be only a partial description
of the internal physics in a Cepheid atmosphere. In particular, at
cooler temperatures as the convective regions become larger and the
dynamical timescale becomes a significant fraction of the pulsation
period our models are expected to become increasingly divergent. This
is a well known shortcoming of models that implement the
mixing-length approximation. Whilst our models can match the blue edge
of the instability strip (IS) they can not reproduce its red edge. In
the vicinity of the red edge the amplitude of pulsation is too
high, a feature that can not be circumvented by modification of
artificial viscosity parameters. To produce a physical red edge an
additional form of energy dissipation is required. Convective
processes are the likely cause of this. \citet{yec98}
shows that models with a parametrised formulation of turbulent
convective mixing are able to match the fundamental and first overtone
instability strips through a fine tuning of parameters.

\begin{figure}
\plotone{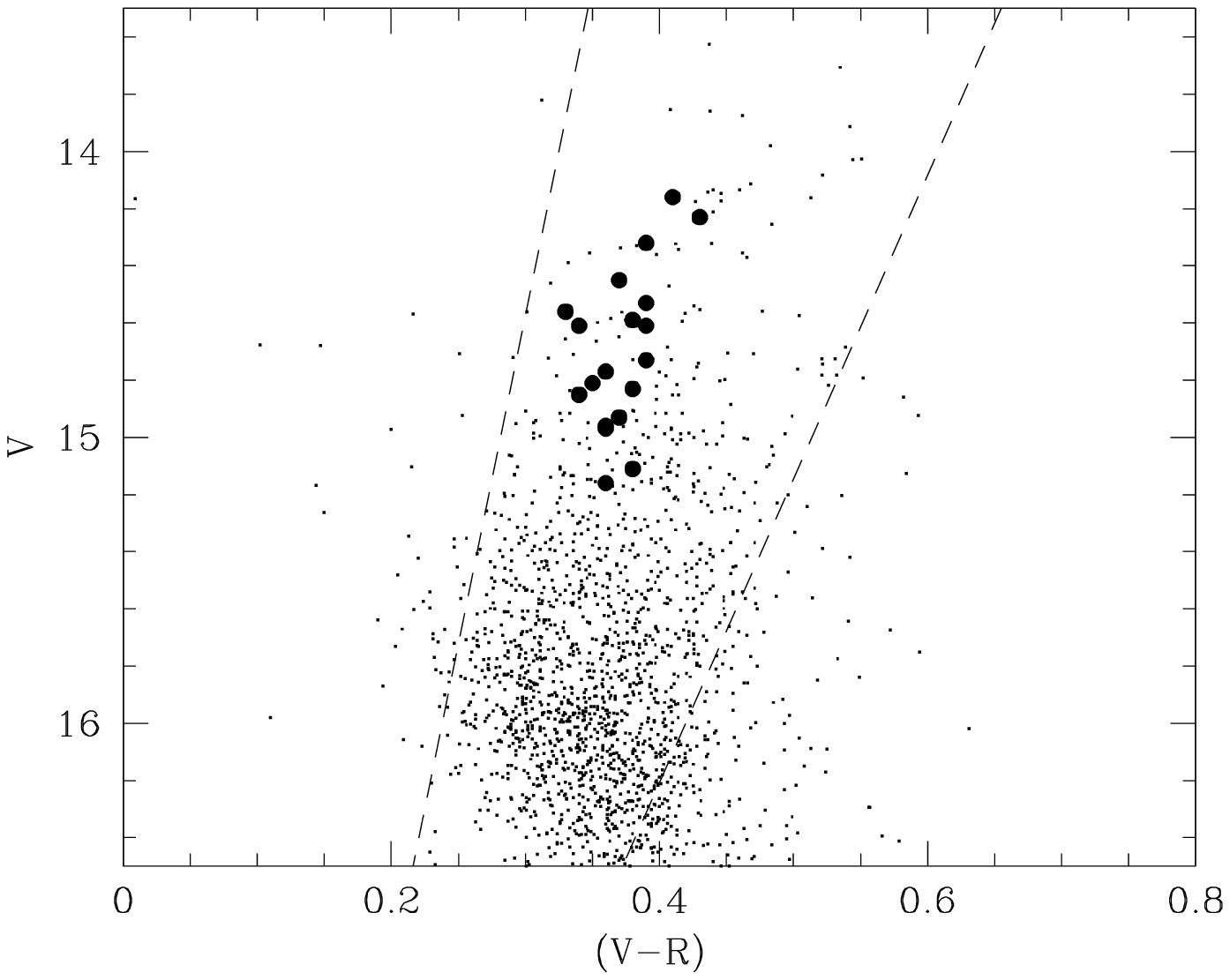}
\caption{The $V$, $V$$-$$R$ colour-magnitude diagram for fundamental
Cepheids in the MACHO catalogue. The bump Cepheids of the present
study are shown in bold. Overlaid are the blue and red edge of the
instability strip described by Chiosi, Wood \& Capitano (1993)
assuming E($B$$-$$V$)=0.12 \label{sample}}
\end{figure}

Yecko et al.\ consider the case of a 5M$_{\odot}$ star modeled both
with mixing-length approximation and with turbulent
convection. Consideration of the model growth rates (their figure 11)
shows insignificant differences over the bluest 1/4 of the IS,
becoming increasingly divergent thereafter. In order to avoid as much
as possible the short comings of the mixing-length approach we have
sought bump Cepheids close to the blue edge of the IS (see
Fig.~\ref{sample}).

\section{Results}

\begin{figure}
\plotone{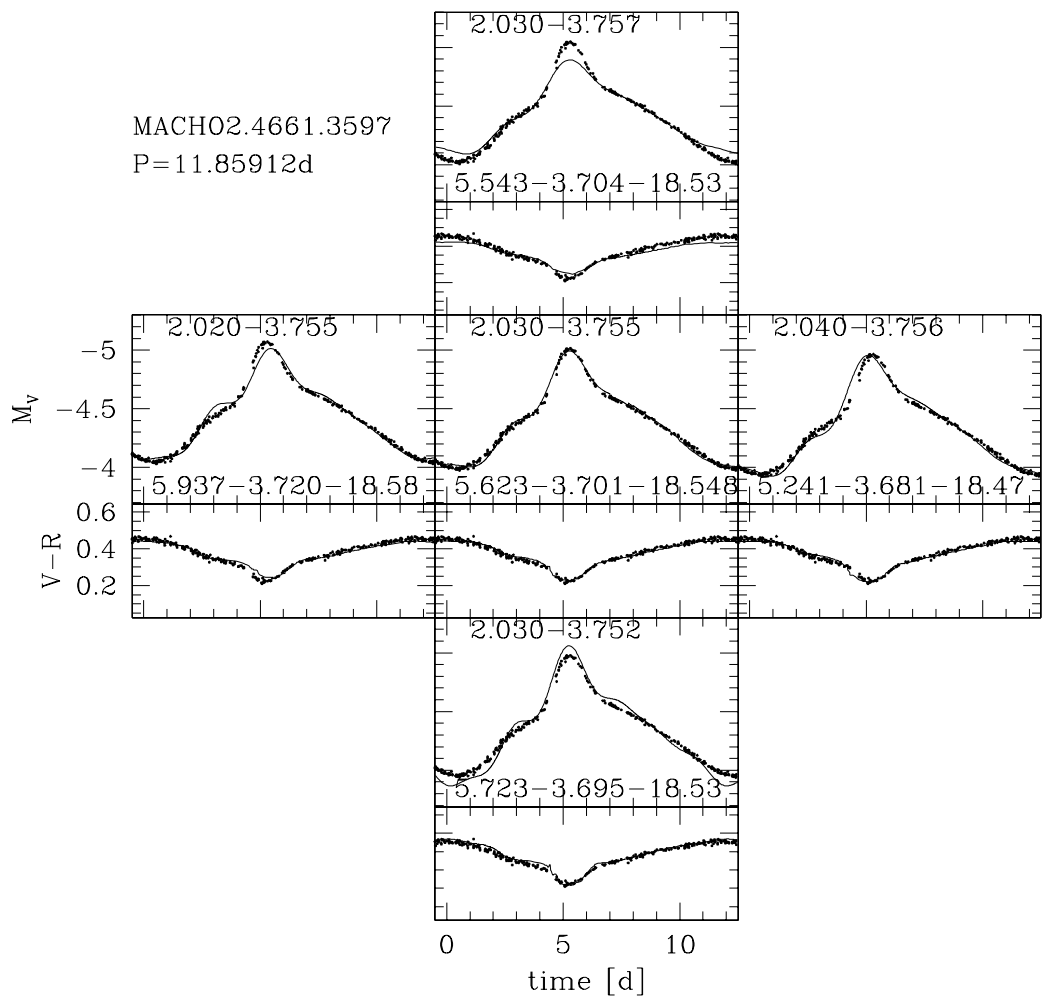}
\caption{Model fits for MACHO~2.4661.3597. The numbers in each panel are: the period ratio P$_{02}$ and log T$_{eff}$ in the upper part, and M/M$_{\odot}$, log L/L$_{\odot}$, and the true distance modulus in the lower part. Each panel contains light and color curves (dereddened). Observations are shown as dots and models as lines. The panels in the vertical section show the effect of changing T$_{eff}$, this affects the amplitude (in the upper panel T$_{eff}$ is too large, the resulting amplitude is too low). The horizontal section shows the effect of a changing P$_{02}$ this affects the phase of the bump (in the left panel P$_{02}$ is too small, the phase of the bump is too ``late''). The central panel is the best model.}\label{figstar1}
\end{figure}

\begin{figure}
\plotone{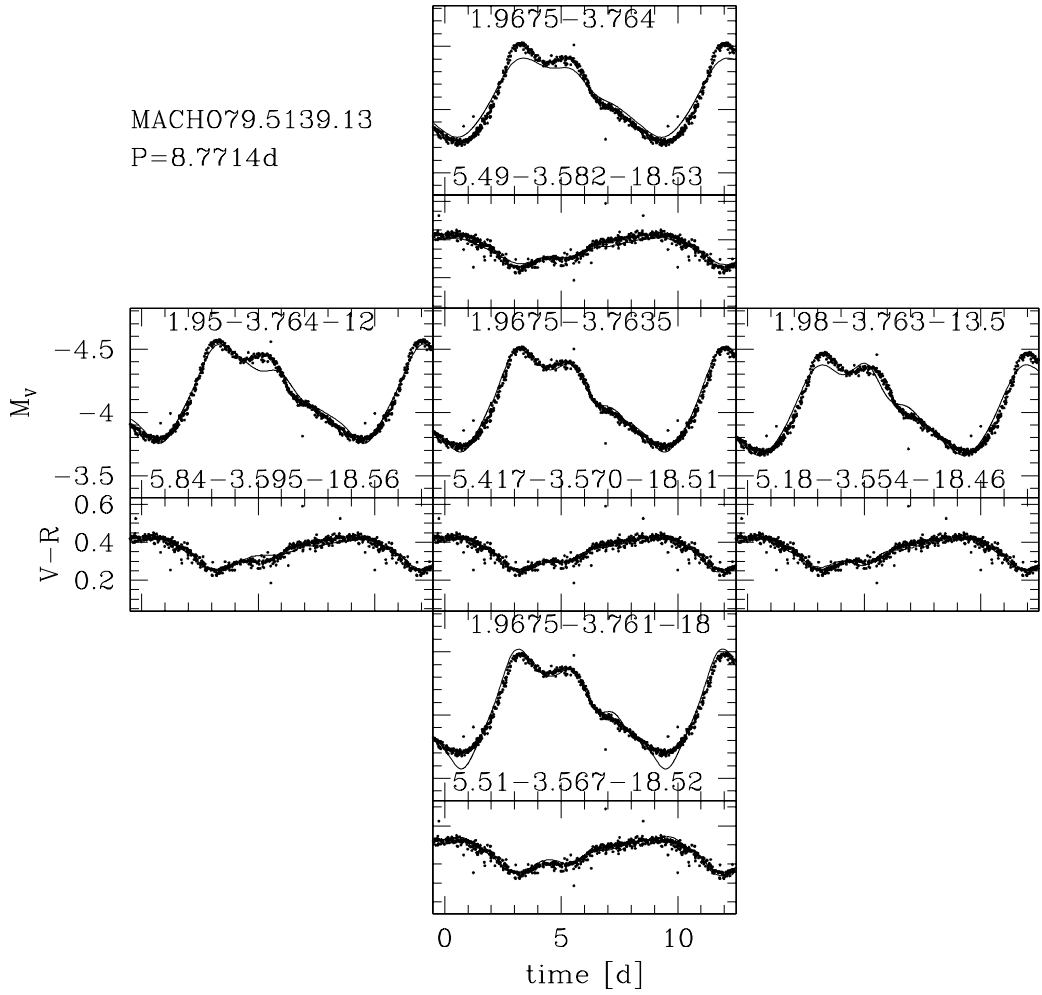}
\caption{As in Fig.~\ref{figstar1}, model fits for MACHO~79.5139.13.\label{figstar2}}
\end{figure}

In figures \ref{figstar1} \& \ref{figstar2} we show the effects of
variation of the two parameters $T_{eff}$ and P$_{02}$. As we change
P$_{02}$ we modify the phase at which the bump is located. Similarly,
as $T_{eff}$ is varied the amplitude is changed. The best fit to the
observed light curve is shown in the central panel. Here for the
purpose of illustration we show five models widely separated in
parameter space. In the determination of the best model, however, we
use a iterative chi-squared minimisation technique.

The offset of the model M$_{V}$ and observed $V$ light curves gives the
apparent distance modulus. Having obtained a model that matches the
$V$ light curve of each bump Cepheid, the observed $V$-$R$ colour
curve was shifted onto the model intrinsic $V$-$R$ colour curve. The shift
required to do so is the colour excess, E$_{V-R}$, which can be
converted to visual absorption, A$_{V}$, using a standard reddening
curve (A$_{V}$=1.78E(${V-R}$) was used). The true distance modulus can
then be obtained from the apparent distance modulus. The best fit
parameters for the Cepheids in our sample are given in Table
\ref{tbl-2}.

Limits of {\emph{internal}} accuracy in the determination of the best
model solution arises from cycle-to-cycle variations in the model
output. By monitoring on the order of 30 cycles we can quantify the
uncertainty introduced by these model variations. The errors in the
determined parameters are dominated by the uncertainty in
P$_{02}$. This becomes problematic towards lower masses as the bump
amplitude diminishes. {\emph{Systematic}} uncertainties dominate the
total uncertainty however. The quoted photometric uncertainties for
the MACHO photometry account for $\pm$0.1M$_{\odot}$ in mass and
$\pm$0.002 dex in luminosity. Error bars shown in
figure~\ref{mlrelfig} are the quadrature sum of systematic and
internal uncertainties.

As stated above, a value of metallicity is adopted in our analysis. It
is important to remark on the effects that varying this metallicity
has on our models. An exploration of this has been presented by us in
Paper 1. In this work, models were additionally constructed for solar
and SMC metallicities. It was found that the models cannot be used to
distinguish the abundance of the Cepheid with any meaningful
uncertainty. If the abundance of our sample is assumed to lie in the
range Z=0.006-0.01 and Y=0.25-0.29 the resulting error in the derived
parameters from the uncertainty due to abundance is of similar
magnitude as that reported from P$_{02}$ and T$_{eff}$.

A less constrainable systematic effect may arise from our treatment of
convection. We note the work of \citet{buc96} and \citet{feu00} who
find that the effects of convection are important even in the vicinity
of the blue edge. Whilst a large degree of freedom is available in the
internal parameters of turbulent convection models the magnitude of
this possible systematic effect remains uncertain.

\subsection{Reddenings and Distance Modulus}

Another independent prediction of the bump Cepheid models is the
reddening to each object. In figure \ref{reddmfig} we present the
histogram of determined reddenings. Our derived mean reddening is
0.08.

A number of studies of line-of-sight redddening to LMC stars have been
presented in the literature. \citet{bes91} showed that the
historically low values for reddening to the LMC
($<E$($B$$-$$V$)$>\sim$0.03) were too low and used the Johnson \&
Morgan reddening-free $Q$ index and spectroscopic temperatures to show
that the $<$E($B$$-$$V$)$>\sim$0.12. \citet{har97}
use $Q$ to form a histogram of E($B-V$) from a 2.8\arcdeg$^2$ region of the
LMC. They find a mean reddening of 0.20 with a non-gaussian tail
extending to higher E($B-V$). \citet{lar00} report
$<$E($B$$-$$V$)$>=0.085$. \citet{zar99} revisits the data of Harris et
al.\ and reports a strong dependence of $<$E($B$$-$$V$)$>$ on the spectral
type of objects used to define it. Namely, low extinctions result from
red clump stars, high extinction from OB stars. Zaritsky proposes that
is the result of a larger scale height for older stars, placing them
statistically in lower reddening regions than the OB stars that
reside in the dusty disk. With this background we find that our
reddenings are applicable for objects within the LMC. 

The mean LMC distance modulus shown in Figure \ref{reddmfig} is
18.55$\pm$0.02. This in good agreement with recent analysis
(18.57$\pm$0.14 from the compilation of Gibson 2000).

The consistency of our derived reddening and distance modulus with
previous studies provides a validation to the input of our model. It
means that the results we present here will be consistent with results
derived from linear pulsation models. Linear pulsation models that
utilize the observed colours, reddenings, apparent luminosity and
distance modulus by \citet{seb95} do indeed show pulsation masses
for LMC Cepheids that are similarly lower than evolution masses. This
gives us confidence in the consistency of the linear and non-linear pulsation
theory as well as the bolometric corrections and colour
transformations.

\subsection{The Mass-Luminosity Relation}

\begin{figure}
\plotone{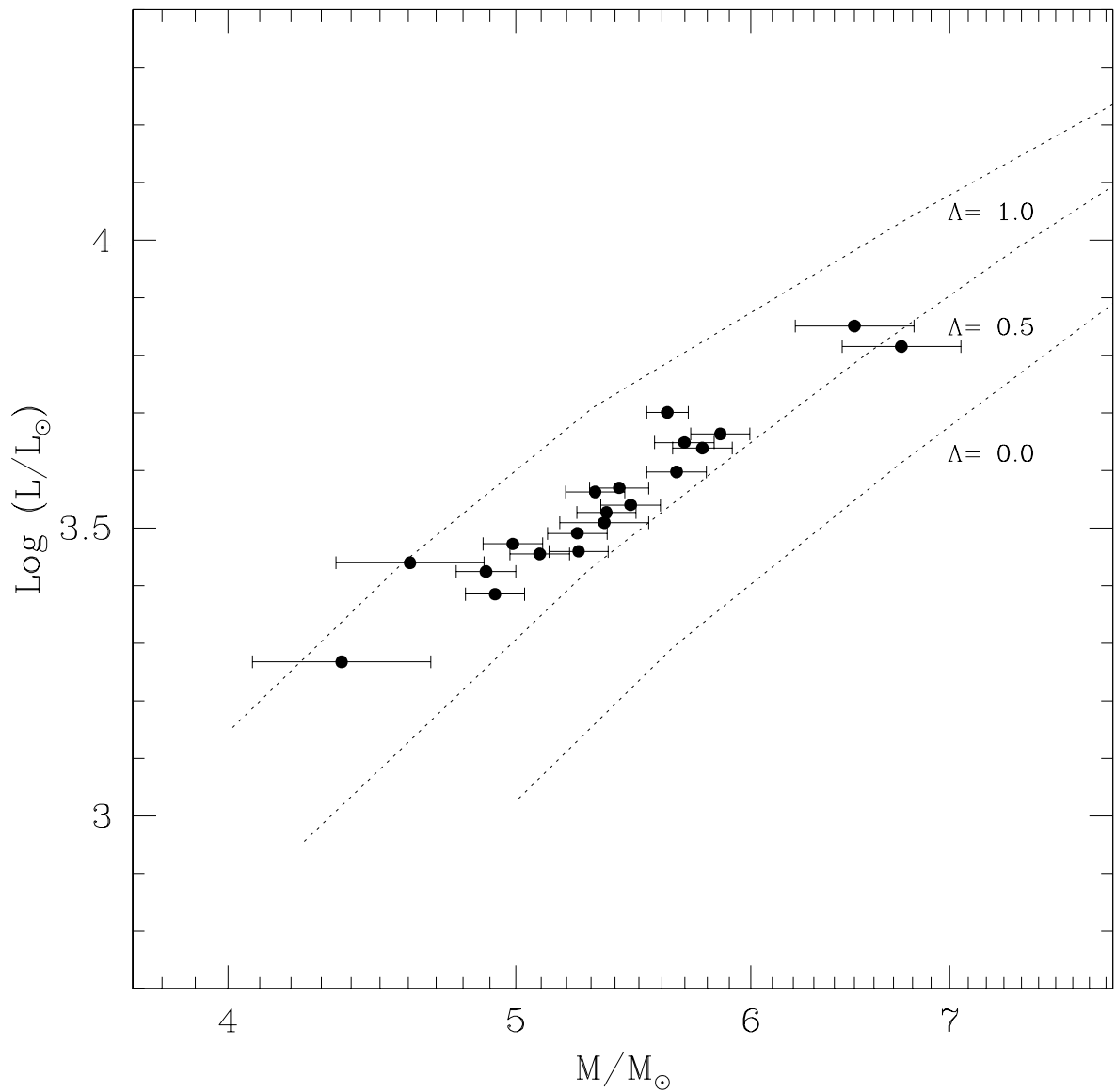}
\caption{The mass-luminosity relation for the present sample of 20
bump Cepheids. Error bars are as discussed in the text. Overlaid are
the M-L relations for core-He burning stars from Fagotto et al.\
(1994) and Bressan (2001) for three assumptions of the efficiency of
convective core overshoot and Z=0.008 and Y=0.25. The solid line
represent the best match to the data ($\Lambda_c$=0.65
$l/H_p$). }\label{mlrelfig}
\end{figure}

\begin{figure}
\plotone{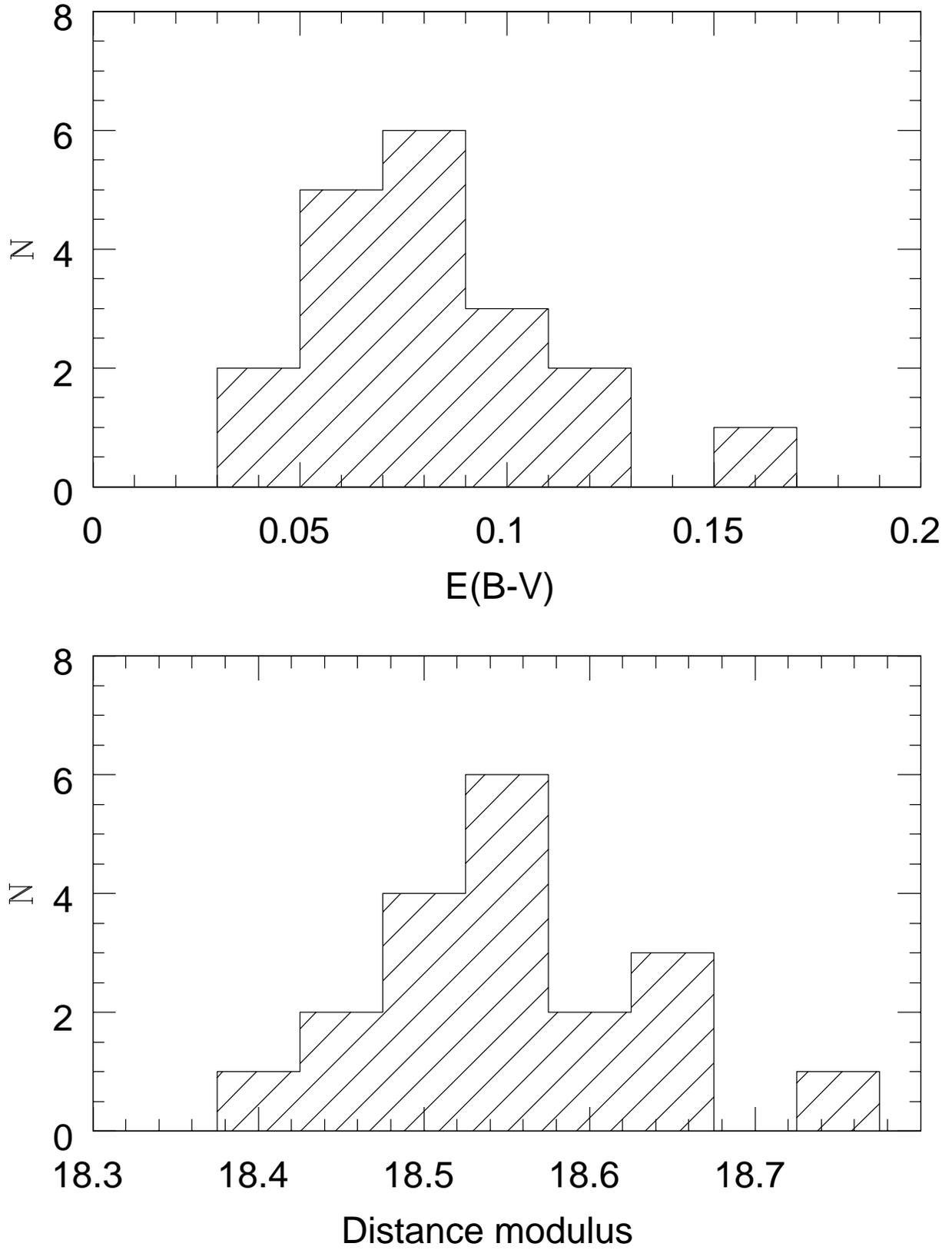}
\caption{Histograms of the derived reddenings (top) and distance moduli (bottom) for the 20 Cepheids of our sample.\label{reddmfig}}
\end{figure}

By modelling the bump Cepheid sample we have independently determined
both $L$ and $M$ for each object. We use these values to construct the
mass-luminosity relation for core-He burning stars in the LMC. Figure
\ref{mlrelfig} shows the mass-luminosity relationship described by our
sample. Overlaid are the M-L relations (due to \citet{fag94} and
\citet{bre01}) for three levels of convective core overshoot
efficiency with Z=0.008 and Y=0.25.

The data are significantly more luminous than the predictions of
standard ``mild'' convective core overshoot
(i.e.~$\Lambda_c=0.5$). Figure \ref{mlrelfig} recommends a degree of
convective core overshoot of $\Lambda_c=0.65\pm0.03 l/H_p$. Put
another way, our results favour a mass 19.5$\pm1.0$\% lower than
classical evolutionary models.

The problem of reconciling mass determinations from the various
techniques available has been a problem that has plagued the
field. The many phases of debate, and their convergence, have been
extensively discussed in the literature (see e.g. \citet{cox80}). The
longest standing of these, the bump and beat Cepheid mass discrepancy,
has to a large part been resolved by \citet{mos92} through the
introduction of OPAL opacities. The discrepancy between pulsation and
evolutionary mass has not been removed by improvements in input
physics.

One possibility that has been suggested by \citet{bon02} is that
mass loss is responsible for the reduction in mass. As opposed to the
evolutionary models used in the study of Bono et al.\ that neglect
mass loss, the models shown in figure \ref{mlrelfig} incorporate the
mass loss prescription of \citet{dej88}. This is largely
responsible for the curvature of the M-L relation towards higher
masses. Furthermore, mass loss during the Cepheid phase does not
appear to be enhanced from that expected from the de Jager description
\citep{dea88}. We conclude that mass loss alone can not explain the Cepheid
mass discrepancy.

One Cepheid (MACHO~2.4661.3597: HV 905) has been the target of
previous analysis in Paper 1. However, the current work uses updated
OPAL opacities and improved low temperature opacities from
\citet{ale94}, and replaces bolometric corrections of \citet{kur93}
with those of \citet{cas99}. The results from this previous work were
$M$=5.15$\pm$0.35M$_{\odot}$ and $log(L/L_{\odot})$=3.69$\pm$0.01
(uncertainties determined post-fact from the details presented in
Paper 1). Here we find $M$=5.62$\pm$0.22M$_{\odot}$ and
$log(L/L_{\odot})$=3.701$\pm$0.004. 

In their description of their non-linear pulsation models that
incorporate turbulent convection, \citet{bon99} compare
their model output for HV 905 with that of Paper 1. They use the mass,
luminosity, T$_{eff}$ and abundance in Paper 1 and retrieve a limit
cycle lightcurve that is very similar in shape to that of Paper 1. In
addition, the P$_{02}$ differs by only 1\% and the luminosity
amplitude differs by 0.06-0.08 mag from that presented in Paper
1. This gives us confidence that our study is unaffected by our
treatment of convection under the mixing length approximation.

Future refinement of the technique described here is possible through
the comparison of observed radial velocity variations of bump Cepheids
to those predicted by our model.  This has the benefit of limiting
systematic errors.  At present, the accuracy of our technique is
limited by systematic photometric uncertainties (in particular
transformations of MACHO bandpasses and zeropoints). The radial
velocity curve is a more robust output of the non-linear pulsation
model. Radial velocities are more closely related to the dynamical
processes within the stellar atmosphere than that of photometry which
is more affected by temperature changes than dynamical effects. Work
is currently underway by us to use radial velocity curves for bump
Cepheids as a stringent test on our pulsation models. Models should be
able to accurately reproduce all observational constraints (period,
shapes of light \emph{and} velocity curves). Such a critical
examination of the dynamics of Cepheid pulsation offers the potential
of further model refinement.

We note that demonstrating the convective core is of greater extent
than the core defined by the Schwarzschild criterion does not
determine its causation. Recent studies have shown that rotation can provide a
natural way to bring about increased internal mixing and hence a
larger core \citep{heg00,mey00}.

Attention to rotational mixing has been promoted by the findings of
chemical abundance studies of A supergiants \citep{ven95}, B
supergiants \citep{duf00}, and main-sequence B stars
\citep{daf01,gie92} which imply a broad range of enhancements. This
behavior is interpreted as arising from self-contamination with
CN-cycled material prior to the first dredge-up. Such surface
modifications can not be achieved through CCO.

Furthermore, the CNO abundances for a set of 10 SMC A supergiants have
been studied by \citet{ven99}. These stars show a large range in N
abundance enhancements ranging from negligible to levels greater than first
dredge-up. The magnitude of enhancement is much greater than that
found for a similar sample of Galactic A supergiants \citep{ven95} which
suggests a possible metallicity effect in the underlying physics of
internal mixing.

Bump Cepheids of the SMC offer the opportunity to investigate the
metallicity dependence of internal mixing. A greater level of internal
mixing such as would be produced by generally higher stellar rotation
velocities in the SMC \citep{ven99} should result in a clear
modification of the M-L relation. Work is underway to look for such effects.

\section{Conclusions}

In this study we have applied non-linear pulsation models to match the
observed light and colour curves of 20 LMC bump Cepheids. We have been
able to place tight constraints on the stellar parameters mass,
luminosity and effective temperature as well as individual
reddenings. We have described the mass-luminosity relation over the
mass range of bump Cepheids. We find our sample is $\sim$20\% more
luminous for their mass predicted by stellar evolution models that do
not incorporate extension to the convective core. Under the paradigm
of convective core overshoot this amounts to $\Lambda_c$ of
0.65$\pm$0.03$l/H_p$. The models yield an LMC distance modulus of
18.55$\pm$0.02.

\acknowledgments

We thank A.\ Bressan et al.\ for providing us with unpublished
evolutionary models for $\Lambda_c$=1.0. We thank our referee
Dr.~Z.~Kollath for his comments and personal calculations regarding
systematic effects due to turbulent convection. Work performed by SCK
was performed under the auspices of the U.S. Department of Energy,
National Nuclear Security Administration by the University of California,
Lawrence Livermore National Laboratory under contract No. W-7405-Eng-47.

\clearpage


\begin{table}
\begin{center}
\caption{The selected MACHO bump Cepheid sample\label{tbl-1}}
\begin{tabular}{lccccl}
\tableline\tableline
{\it{MACHO}} star id & RA (J2000) & Dec (J2000) & $<V>$ & $<V$$-$$R>$ & $P$[d] \\
\tableline

1.3441.15 & 05 01 52.0 & -69 23 22 & 14.45 &0.37& 10.4136\\
1.3692.17 & 05 02 51.4 & -68 47 06 & 14.53 &0.39&10.8552\\
1.3812.15 & 05 03 57.3 & -68 50 25 & 14.61 & 0.39&9.7118\\
1.4048.6  & 05 05 08.8 & -69 15 12 & 14.77 & 0.36& 7.7070\\
1.4054.15 & 05 05 42.1 & -68 51 06 & 14.93 &0.37 & 7.3953 \\
2.4661.3597 & 05 09 16.0 & -68 44 30 & 14.32& 0.39& 11.85911\\
6.6456.4346 & 05 20 23.1 & -70 02 33 & 15.16& 0.36& 6.4816\\
9.4636.3 & 05 09 04.5 & -70 21 55 & 14.16 & 0.41&  13.6315\\
9.5240.10 & 05 13 10.1 & -70 26 47 & 15.11 & 0.38& 7.3695\\
9.5608.11 & 05 15 04.7 & -70 07 11 & 14.81 & 0.35& 7.0693\\
18.2842.11 & 04 57 50.2 & -68 59 23 & 14.83 & 0.38& 8.8311\\
19.4303.317 & 05 06 39.8& -68 25 13 & 14.65 & 0.37& 8.7133\\
19.4792.10 & 05 09 36.9 & -68 02 44 & 14.96 & 0.36& 6.8628\\
77.7670.919 & 05 27 55.1 & -69 48 05 & 14.85&  0.34 & 7.4423\\
77.7189.11 & 05 24 33.3 & -69 36 41 & 14.73 &  0.39& 7.7712\\
78.6581.13 & 05 20 56.0 & -69 48 19 & 14.97 & 0.36& 6.9302\\
79.4657.3939 & 05 08 49.4 & -68 59 59 & 14.23 & 0.43& 13.8793\\
79.4778.9 & 05 09 56.3 & -68 59 41 & 14.56 & 0.33& 8.1868\\
79.5139.13 & 05 11 53.1 & -69 06 49 & 14.59 &0.38&  8.7716\\
79.5143.16 & 05 12 18.8 & -68 52 46 & 14.61 &0.34&  8.2105\\
\tableline
\end{tabular}
\end{center}
\end{table}

\begin{table}
\begin{center}
\caption{The details of our sample\label{tbl-2}}
\begin{tabular}{lrcccccc}
\tableline\tableline
{\it{MACHO}} star id & $P$[d] & P$_{02}$ & log($T_{eff}$) & E($B-V$) & $\mu_{obj}$ & log(L/L$_{\odot}$) & M/M$_{\odot}$\\
\tableline

1.3441.15 & 10.4136& 2.005& 3.757& 0.078 & 18.554 & 3.649 & 5.698\\
1.3692.17 & 10.8552& 2.012& 3.756& 0.072 & 18.622 & 3.653 & 5.793 \\
1.3812.15 & 9.7118& 1.980& 3.762& 0.076 & 18.569 & 3.639 & 5.778\\
1.4048.6  & 7.7070& 1.938& 3.765& 0.060 & 18.626 & 3.455 & 5.093\\
1.4054.15 & 7.3953& 1.940& 3.767& 0.065 & 18.563 & 3.472 & 4.988\\
2.4661.3597 & 11.8591& 2.030& 3.755& 0.088 & 18.548 & 3.701 & 5.623\\
6.6456.4346 & 6.4816& 1.940& 3.770& 0.093 & 18.665 & 3.267 & 4.368\\
9.4636.3 &  13.6315& 2.035& 3.756& 0.078 & 18.478 & 3.851 & 6.501\\
9.5240.10 & 7.3695& 1.952& 3.767& 0.082 & 18.477 & 3.439 & 4.606\\
9.5608.11 & 7.0693& 1.92& 3.765& 0.155 & 18.735 & 3.460 & 5.249\\
18.2842.11 & 8.8311& 1.973& 3.762& 0.049 & 18.431 & 3.563 & 5.317\\
19.4792.10 & 6.8628& 1.924& 3.766& 0.051 & 18.548 & 3.424 & 4.886\\
19.4303.317 & 8.7133& 1.958& 3.763& 0.068 & 18.564 & 3.597 & 5.663\\
77.7670.919 & 7.4423& 1.936& 3.765& 0.078 & 18.543 & 3.491 & 5.244  \\
77.7189.11 & 7.7712& 1.940& 3.764& 0.092 & 18.501 & 3.509 & 5.355 \\
78.6581.13 & 6.9302& 1.930& 3.764& 0.084 & 18.540 & 3.385& 4.920\\
79.4657.3939 & 13.8793& 2.032& 3.758& 0.112 &  18.528 & 3.815& 6.742\\
79.4778.9 &  8.1868& 1.953& 3.762& 0.071 & 18.643 & 3.527& 5.364\\
79.5139.13 &  8.7716& 1.968& 3.763& 0.068 & 18.509 & 3.570 & 5.417\\
79.5143.16 &  8.2105& 1.95& 3.763& 0.114 & 18.571 & 3.540 & 5.465\\
\tableline
\end{tabular}
\end{center}
\end{table}


\begin{thebibliography}{}
\bibitem[Alexander \& Ferguson(1994)]{ale94} Alexander, D. R. \& Ferguson, J. W. 1994, \apj, 437, 879
\bibitem[Alcock et al.(1999)]{alc99} Alcock, C.~et al.\ 1999, 
\pasp, 111, 1539
\bibitem[Barmina et al.(2002)]{bar02} Barmina, R., Girardi, L.,Chiosi, C.\ 2002, \aap accepted, astro-ph/0202128 
\bibitem[Beaulieu et al.(2001)]{bea01} Beaulieu, J.P., Buchler, J.R. \& Kollath, Z.\ 2001, A\&A, 373, 164
\bibitem[Bessell(1990)]{bes90} Bessell, M.~S.\ 1990, \pasp, 102, 1181
\bibitem[Bessell(1991)]{bes91} Bessell, M.~S.\ 1991, \aap, 242, L17
\bibitem[Bessell, Castelli, \& Plez(1998)]{bes98} Bessell, M.~S., Castelli, F., \& Plez, B.\ 1998, \aap, 333, 231
\bibitem[Bohm-Vitense et al.(1997b)]{boh97b} B\"ohm-Vitense, E., Evans, N.~R., Carpenter, K., Morgan, S., Beck-Winchatz, B., \& Robinson, R.\ 1997b, \aj, 114, 1176. 
\bibitem[Bohm-Vitense et al.(1997a)]{boh97a} B\"ohm-Vitense, E., Remage Evans, N., Carpenter, K., Beck-Winchatz, B., \& Robinson, R.\ 1997a, \apj, 477, 916. 
\bibitem[Bono, Castellani, \& Marconi(2002)]{bon02} Bono, G., Castellani, V., \& Marconi, M.\ 2002, \apjl, 565, L83
\bibitem[Bono, Marconi, \& Stellingwerf(1999)]{bon99} Bono, G., Marconi, M., \& Stellingwerf, R.~F.\ 1999, \apjs, 122, 167
\bibitem[Bressan, Chiosi, \& Bertelli(1981)]{bre81} Bressan, A.~G., Chiosi, C., \& Bertelli, G.\ 1981, \aap, 102, 25
\bibitem[Bressan(2001)]{bre01} Bressan, A.~G., 2001, priv.\ comm.
\bibitem[Buchler et al.(1996)]{buc96} Buchler, J.~R., Kollath, Z., Beaulieu, J.-P. \& Goupil, M.~J.,  ApJ, 462, L83
\bibitem[Castelli(1999)]{cas99} Castelli, F. 1999, A\&A 346, 564
\bibitem[Chiosi et al.(1993)]{chi93} Chiosi, C., Wood, P.R., \& Capitanio, N. 1993, A\&AS, 86, 541
\bibitem[Cordier et al.(2002)]{cor02} Cordier, D., Lebreton, Y., Goupil, M.-J., Lejeune, T., Beaulieu, J.-P. \& Arenou, F.\ 2002, A\&A submitted
\bibitem[Cox(1980)]{cox80} Cox, A.~N.\ 1980, \araa, 18, 15
\bibitem[Daflon et al.(2001)]{daf01} Daflon, S., Cunha, K., Butler, K., \& Smith, V.~V.\ 2001, \apj, 563, 325
\bibitem[de Jager et al.(1988)]{dej88} de Jager, C., Nieuwenhuijzen, H., \& van der Hucht, K.~A.\ 1988, \aaps, 72, 259 
\bibitem[Deasy(1988)]{dea88} Deasy, H.~P.\ 1988, \mnras, 231, 
\bibitem[Dufton et al.(2000)]{duf00} Dufton, P.~L., McErlean, N.~D., Lennon, D.~J., \& Ryans, R.~S.~I.\ 2000, \aap, 353, 311
\bibitem[Evans et al.(1998)]{eva98} Evans, N.~R., B\"ohm-Vitense, E., Carpenter, K., Beck-Winchatz, B., \& Robinson, R.\ 1998, \apj, 494, 768
\bibitem[Evans et al.(1997)]{eva97} Evans, N.~R., B\"ohm-Vitense, E., Carpenter, K., Beck-Winchatz, B., \& Robinson, R.\ 1997, \pasp, 109, 789
\bibitem[Fagotto et al.(1994)]{fag94} Fagotto, F., Bressan, A., Bertelli, G., \& Chiosi, C.\ 1994, \aaps, 105, 29
\bibitem[Feuchtinger, Buchler \& Kollath(2000)]{feu00} Feuchtinger, M., Buchler, J.~R. \& Kollath, Z.\ 2000, \apj, 544, 1056 
\bibitem[Freedman et al.(2001)]{fre01} Freedman, W.~L.~et al.\ 2001, \apj, 553, 47
\bibitem[Gibson(2000)]{gib00} Gibson, B.\ 2000, Mem.\ Ast.\ It., astro-ph/9910574
\bibitem[Gies \& Lambert(1992)]{gie92} Gies, D.~R.~\& Lambert, D.~L.\ 1992, \apj, 387, 673
\bibitem[Harris, Zaritsky, \& Thompson(1997)]{har97} Harris, J., Zaritsky, D., \& Thompson, I.\ 1997, \aj, 114, 1933
\bibitem[Heger, Langer, \& Woosley(2000)]{heg00} Heger, A., Langer, N., \& Woosley, S.~E.\ 2000, \apj, 528, 368 
\bibitem[Hertzsprung(1926)]{hea26} Hertzsprung, E. 1926, Bull. astronom. Inst. Netherl., 3, 115
\bibitem[Iglesias \& Rogers(1996)]{igl96} Iglesias, C.A. \& Rogers, F.J. 2996, \apj, 464, 943 
\bibitem[Kanbur \& Simon(1994)]{kan94} Kanbur, S.~M.~\& Simon, N.~R.\ 1994, \apj, 420, 880
\bibitem[Keller, Da Costa, \& Bessell(2001)]{kel01} Keller, S.~C., Da Costa, G.~S., \& Bessell, M.~S.\ 2001, \aj, 121, 905 
\bibitem[Kurucz(1993)]{kur93} Kurucz, R.L.\ 1993, CD-ROM No.~13.\  Cambridge, Mass.: Smithsonian Astrophysical Observatory
\bibitem[Larsen, Clausen, \& Storm(2000)]{lar00} Larsen, 
S.~S., Clausen, J.~V., \& Storm, J.\ 2000, \aap, 364, 455
\bibitem[Meynet \& Maeder(2000)]{mey00} Meynet, G.~\& Maeder, A.\ 2000, \aap, 361, 101 
\bibitem[Moskalik, Buchler, \& Marom(1992)]{mos92} Moskalik, P., Buchler, J.~R., \& Marom, A.\ 1992, \apj, 385, 685
\bibitem[Russell \& Bessell(1989)]{rus89} Russell, S.~C.~\& Bessell, M.~S.\ 1989, \apjs, 70, 865 
\bibitem[Sebo \& Wood(1995)]{seb95} Sebo, K.~M.~\& Wood, P.~R.\ 1995, \apj, 449, 164
\bibitem[Simon \& Lee(1981)]{sim81} Simon, N.~R.~\& Lee, A.~S.\ 1981, \apj, 248, 291 
\bibitem[Stobie(1969)]{sto69} Stobie, R.~S.\ 1969, \mnras, 144, 511.
\bibitem[Stothers \& Chin(1992)]{sto92} Stothers, R.~B.~\& Chin, C.\ 1992, \apj, 390, 136 
\bibitem[Venn(1999)]{ven99} Venn, K.~A.\ 1999, \apj, 518, 405 
\bibitem[Venn(1995)]{ven95} Venn, K.~A.\ 1995, \apj, 449, 839
\bibitem[Wood, Arnold, \& Sebo(1997)]{woo97} Wood, P.~R., Arnold, A., \& Sebo, K.~M.\ 1997, \apjl, 485, L25 (Paper 1)
\bibitem[Yecko, Kollath, \& Buchler(1998)]{yec98} Yecko, P.~A., Kollath, Z., \& Buchler, J.~R.\ 1998, \aap, 336, 553
\bibitem[Zaritsky(1999)]{zar99} Zaritsky, D.\ 1999, \aj, 118, 
2824 
\end{thebibliography}
\end{document}